\documentclass[nofootibib,preprint,superscriptaddress]{revtex4}
\usepackage{color}
\usepackage{amssymb}
\usepackage{graphicx}
\usepackage[normalem]{ulem} 
\renewcommand\sout{\bgroup\color[rgb]{0,0,1} \ULdepth=-.5ex \ULset}

\begin{document}
\title{Inclusive electron scattering in the quasielastic region with Korea-IBS-Daegu-SKKU density functional}
\author{Hana Gil}
\affiliation{Center for Extreme Nuclear Matter, Korea University, Seoul 02841, Korea}
\author{Chang Ho Hyun}
\affiliation{Department of Physics Education, Daegu University, Gyeongsan 38453, Korea}
\author{Kyungsik Kim}
\email{kyungsik@kau.ac.kr}
\affiliation{School of Liberal Arts and Science, Korea Aerospace University, Goyang 10540, Korea}
\date{\today}
\begin{abstract}
With the framework of KIDS (Korea-IBS-Daegu-SKKU) density functional model, the isoscalar and isovector effective masses of nucleon and the effect of symmetry energy in nuclear medium  are investigated in inclusive $(e,e')$ reaction in quasielastic region.
The effective masses are varied in the range $(0.7 \sim 1.0)M$ with free nucleon mass $M$,
and the symmetry energy is varied within the uncertainty allowed by nuclear data and neutron star observation.
The wave functions of nucleons inside target nucleus are generated by solving Hartree-Fock equation with adjusting equation of state, binding energy and radius of various stable nuclei, and effective mass of nucleon in the KIDS model.
With the obtained wave functions, we calculate the differential cross section for the inclusive $(e,e')$ reaction and compare the theoretical results with Bates, Saclay, and SLAC experimental data.
Our model describes experimental data better at SLAC-type high incident electron energy than those measured from Bates and Saclay.
The influence of the effective mass and symmetry energy appears to be precise on the longitudinal cross section.
\end{abstract}
\maketitle
\section{Introduction}
The electron scattering has long been acknowledged as one of a useful tools in exploring the structure of nucleus and the dynamics of nucleons in nuclear medium.
In a recent work on the exclusive quasielastic scattering of electrons off nuclei from light to heavy nuclei, 
we observed that the cross sections and response functions have systematic dependence on 
the effective mass of the nucleon in nuclear medium \cite{exeA2021}.
In particular, the dependence appears evident in $^{208}$Pb, in which effective mass close to free mass reproduces 
well the experimental data of the single particle levels of the knocked out protons, and their cross sections.

While exclusive $(e,e'p)$ reaction is concerned with the distribution of energy levels close to the Fermi surface,
inclusive $(e,e')$ reaction subsumes the participation of all the nucleons in a nucleus.
The shape and width of the peak in the inclusive cross section depends on the average momentum and the distribution
of energies of nucleons bound in nuclei.
Therefore inclusive $(e,e')$ reaction provides a unique chance to probe microscopic properties of nuclei
not limited to a few specific and individual states, but contributed from the whole nucleons inside a nucleus.

In the viewpoint of theory, inclusive $(e,e')$ reaction provides opportunities to test the performance and validity of models employed
in the description of nuclear structure and modification of hadronic properties in nuclear medium.
Nuclear structure models are good at reproducing the static bulk properties nuclei such as binding energy and charge radius.
Although single particle levels are summed up to produce correct binding energies, their distributions are strongly dependent of the nuclear structure models.
For example, it has been shown that the distribution of single particle levels depends on the effective mass of nucleons in nuclear medium.
Survey over the 240 Skyrme force models \cite{dutra2012} shows that the effective mass of the nucleon varies in the range $(0.58-1.12) M$ 
where $M$ is the free mass of the nucleon.
Since it receives contributions from all the single particle levels in a nucleus, 
inclusive $(e,e')$ reaction can shed light on reducing uncertainties of the effective mass.

Another topic under hot debate is the density dependence on the symmetry energy \cite{se}. 
It is quite certain that the excessive spatial occupation of the neutron in neutron-rich nuclei is strongly 
correlated to the slope of symmetry energy \cite{nskin}.
Since electron scattering is a qualified method to examine the spatial structure of nuclei, inclusive $(e,e')$ reaction provides a non-nucleonic probe
to explore the uncertainty of the density dependence of nucear symmetry energy.
Two dominant uncertainties in the nuclear many-body physics, in-medium effective mass of the nucleon and correct form of the symmetry
energy as a function of density have not been studied systematically in the inclusive electron scattering.

Main interest of the work is twofold:
First, we test a recently developed model for nuclear structure and nuclear matter by calculating the cross section 
in the inclusive electron quasielastic scattering from nuclei and by comparing the model results with experimental data.
We focus on the kinematics available from Bates \cite{bates}, Saclay \cite{saclay}, and SLAC \cite{slacdata} experiments, 
in which target nuclei are $^{12}$C, $^{40}$Ca, $^{56}$Fe, $^{197}$Au, and $^{208}$Pb
and cross sections can be extracted for incident electron energies up to 2.02 GeV and compared with experimental data.
Second, the issue of uncertainties pertinent to the effective mass and symmetry energy is investigated by calculating the scattering
observables with different values of the effective mass and the symmetry energy.
We observed a manifest dependence on the effective mass that is critical to determination of observables from the consideration of 
exclusive $(e,e'p)$ reaction.
For the sake of consistency, and to understand the role of effective mass clearly, we use the effective mass values used in \cite{exeA2021},
i.e. $0.7 M$ and $0.9 M$ for both isoscalar and isovector effective masses.
Density dependence of the symmetry energy has been explored extenstively by using both nuclear and neutron star data \cite{nse, k0}.
In Ref. \cite{k0}, compression modulus of the symmetric nuclear matter and symmetry energy parameters are constrained, and the final
results are summarized in 4 sets, in which the slope parameter varies in the range 45 $\sim$ 65 MeV.
In order to highlight the dependence on the value of slope parameter, we employ model KIDS-A and KIDS-C in this work.

The paper is organized in the following order.
Section II introduces the model,
and Section III presents the basic formalism for the scattering cross section. 
Section IV shows the results and discussion, and we summarize the work in Section V.

\section{Model}

A great advantage of the KIDS formalism is that the equation of state (EoS) of infinite nuclear matter,
bulk properties of nuclei, and dynamical properties such as in-medium effective mass of the nucleon can be treated independently to each other.
For example, it's been shown that the effective mass of the nucleon can take different values
even if the binding energy and charge radius of nuclei are reproduced at a similar accuracy \cite{kids-nuclei1}.
In Ref.~\cite{exeA2021}, the role of the effective mass is investigated in the exclusive $(e,e')$ reaction in the quasielastic region.
Considered models predict the cross section agreeing well with experiment on the average.
However, detailed comparison shows that the cross section depends on the effective mass in a well defined manner:
models with similar effective masses predict very similar cross sections,
and the results with different effective masses can be clearly distinguished from one another.
Large effective mass models predict results agreeing to data better than the small effective mass models.

In the exclusive $(e,e'p)$ reaction nucleons participating in the process are concentrated to the energy levels and orbital states close to the Fermi surface.
In the inclusive $(e,e')$ process, on the other hand, the contribution of all the nucleons in a nucleus is accounted,
so it provides a more thorough examination of the nuclear structure and nuclear models.
In this work we probe the uncertainty due to the effective mass of nucleons, and density dependence of the nuclear symmetry energy.
Isoscalar and isovector effective mass ratios are defined as
\begin{eqnarray}
\mu_S = \frac{m^*_S}{M}, \,\,\, 
\mu_V = \frac{m^*_V}{M}, 
\end{eqnarray}
where $m^*_S$ and $m^*_V$ are the isoscalar and isovector effective masses, respectively.
With respect to the effective mass, we use three models KIDS0, KIDS0-m*77 and KIDS0-m*99 in Ref. \cite{exeA2021}.
As one can see in Table I, models labeled KIDS0 share the same nuclear EoS specified by $K_0$, $J$, $L$, $K_{\rm sym}$, 
and they differ only in the effective mass.
The other sets of models named KIDS-A and KIDS-C which are rooted in Ref. \cite{k0} explore the effect of nuclear matter properties.
KIDS0, KIDS-A and KIDS-C models have different values of the compression modulus, and symmetry energy parameters $J$, $L$ and $K_{\rm sym}$
which are defined in the conventional expansion of the symmetry energy $S(\rho)$ around the saturation density $\rho_0$
\[
S(\rho) = J + L x + \frac{1}{2} K_{\rm sym} x^2 + \cdots
\]
with $x = (\rho - \rho_0)/3 \rho_0$.
Since there is no adjustment of the effective mass in the KIDS0, KIDS-A and KIDS-C models,
the models have similar values for the effective mass.

\begin{table}[t]
\begin{center}
\begin{tabular}{lcccccc} \hline
Model & $\mu_S$ & $\mu_V$ & $K_0$ &  $J$ & $L$ & $K_{\rm sym}$ \\ \hline
KIDS0 & 0.99 & 0.81 & 240 & 33 & 49 & $-156.2$ \\
KIDS0-m*77 & 0.70 & 0.70 & 240 & 33 & 49 & $-156.2$ \\
KIDS0-m*99 & 0.90 & 0.90 & 240 & 33 & 49 & $-156.2$ \\
KIDS-A & 1.01 & 0.81 & 230 & 33 & 66 & $-139.5$ \\
KIDS-C & 0.98 & 0.80 & 250 & 31 & 58 & $-91.5$ \\
SLy4 & 0.70 & 0.80 & 230 & 32 & 48 & $-119.7$ \\
\hline
\end{tabular}
\end{center}
\caption{Effective mass, compression modulus $K_0$ values and the symmetry energy coefficients of the  considered models.
Dimension of $K_0$, $J$, $L$ and $K_{\rm sym}$ is MeV.}
\label{tab1}
\end{table}

Nuclear contribution to the cross section is evaluated with the initial bound and final scattering states.
In order to preserve the gauge invarinace (i.e. orthonormality of the initial and final state wave functions), 
the same scalar and vector nuclear potential for the initial bound and final nucleon wave functions should be used.
For the nucleon wave functions, we start from the Dirac equation
\begin{eqnarray}
\left[ \gamma_\mu p^\mu - M - S(r) - \gamma^0 V(r) \right] {\bf \Psi} ({\bf r}) = 0,
\end{eqnarray}
where $S(r)$ is the scalar potential, and $V(r)$ is the time component of the vector potential.
We assume spherical symmetry.
Redefining the wave function as
\begin{equation}
{\bf \Psi}'({\bf r}) = {\rm exp}\left[ -\frac{1}{2} \gamma^0 \ln D(r) \right] {\bf \Psi}({\bf r})
\end{equation}
where $D(r)$ is the Darwin non-locality factor, Dirac equation is rewritten as
\begin{equation}
\left[ \gamma_\mu p^\mu -M - \frac{1}{2} {\cal U}(r) - \frac{1}{2}\gamma^0 {\cal U}(r)
+ i {\bf \alpha} \cdot \hat{\bf r} \, T(r) \right] {\bf \Psi}'({\bf r})=0,
\end{equation}
where
\begin{eqnarray}
{\cal U} &=& S + V + \frac{M -E + S + V}{M + E} (S-V), \\
T &=& -\frac{1}{2} \frac{d}{dr} \ln D(r).
\end{eqnarray}
Since the KIDS formalism is based on non-relativistic phenomenology, the potentials we obtain as a result
of the solution of Hartree-Fock equation are non-relativistic potentials.
We have to obtain relativistic potentials $S(r)$ and $V(r)$ from the obtained non-relativistic ones.
Given the non-relativistic central potential $V_{\rm cen}(r)$, Coulomb $V_{\rm Coul}(r)$,
and spin-orbit one $V_{\rm so}(r)$, we calculate the relativistic potentials from relations given in \cite{sv2021}
\begin{eqnarray}
S(r) &=& \frac{1}{2} \left\{ \frac{1}{D} \left[ U - (M-E)(D-1) \right] + (M+E)(D-1) \right\}, \\
V(r) &=& \frac{1}{2} \left\{ \frac{1}{D} \left[ U - (M-E)(D-1) \right] - (M+E)(D-1) \right\},
\end{eqnarray}
where
\begin{eqnarray}
D(r) &=& {\rm exp}\left[-2 \int^\infty_r M r V_{\rm so}(r) dr \right], \\
U(r) &=& U_{\rm nc}(r) + U_{\rm c}(r), \nonumber \\
U_{\rm nc}(r) &=& \frac{2M}{E+M} \left[ V_{\rm cen} + \frac{1}{2M} T^2 + \frac{2 T}{r} + \frac{dT}{dr} \right], \\
U_{\rm c}(r) &=& \frac{2M}{E+M} V_{\rm Coul}, \\
T(r) &=& M r V_{\rm so}.
\end{eqnarray}

\begin{figure}
\begin{center}
\includegraphics[width=0.85\textwidth]{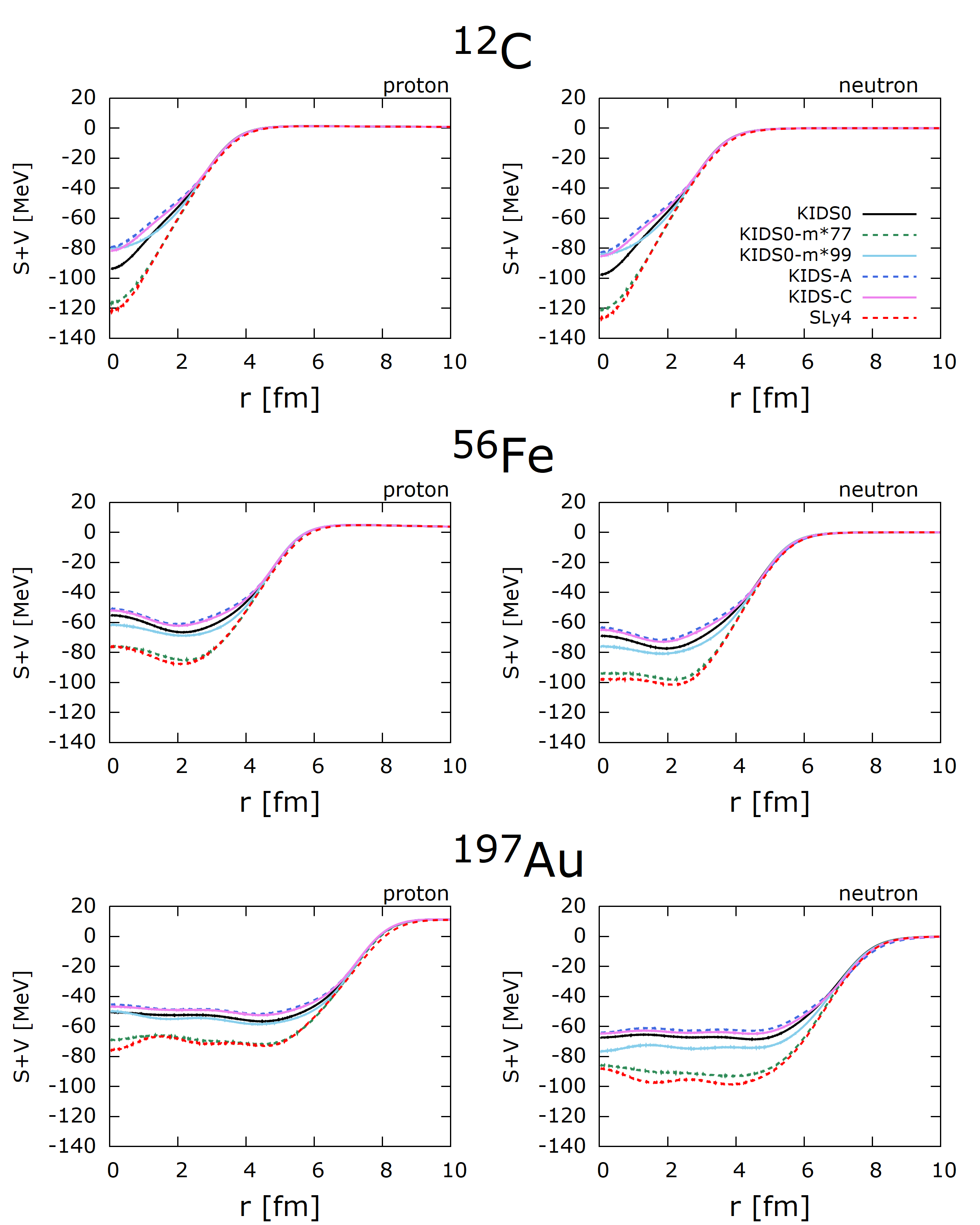} \\
\end{center}
\caption{$S+V$ as a function of $r$ for the denoted nuclei.
Line types noted in the panel of $^{12}$C apply to all the curves in $^{56}$Fe and $^{197}$Au.}
\label{fig:sv}
\end{figure}

Figure \ref{fig:sv} shows the resulting $S(r)+V(r)$ for light ($^{12}$C), medium ($^{56}$Fe), 
and heavy ($^{197}$Au) nuclei with models considered in this work.
In all the cases, KIDS0-m*77 and SLy4 are very similar, and their potentials are always deeper than the other 4 models
KIDS0, KIDS0-m*99, KIDS-A and KIDS-C.
Interestingly, isoscalar effective mass of KIDS-m*77 and SLy4 are the same, and it is similar in the other 4 models.

\section{Formalism}

In the plane wave Born approximation in which the electrons are described as Dirac plane wave, 
the inclusive $(e,e')$ cross section in the rest frame of target nucleus can be written as
\begin{equation}
{\frac {d^2 \sigma} {d\omega d\Omega_e} } = \sigma_M  \left [ {\frac
{Q^4} {q^4}} S_L (q,\omega) + \left (\tan^2 {\frac {\theta_e} {2}} +
{\frac {Q^2} {2q^2}} \right )S_T (q,\omega) \right ], \label{csr}
\end{equation}
where $Q^2 = {\bf q}^2 - \omega^2$ is the four momentum transfer square, $\sigma_M$ represents the Mott cross section given by $\sigma_M = \left ({\frac {\alpha} {2E}} \right )^2 \cos^2 ({\frac {\theta_e}{2}}) / \sin^4 ({\frac {\theta_e}{2}})$, and $S_L$ and $S_T$ are the longitudinal and transverse structure functions which depend only on the three momentum transfer $q (=|{\bf q}|)$ and the energy transfer $\omega$.
By keeping the three momentum and energy transfers fixed while varying the electron incident energy $E$ and scattering angle $\theta_e$, it is experimentally possible to extract the two structure functions with two measurements.

The longitudinal and transverse structure functions in Eq. (\ref{csr}) are squares of the Fourier transform of the components of the nuclear transition current density integrated over outgoing
nucleon angles, $\Omega_p$.
Explicitly, the structure functions for a given bound state with total angular momentum $j_{b}$ are given by
\begin{eqnarray}
S_{L}(q,{\omega})&=&\sum_{{\mu}_{b},\, s_{p}}{\frac {{\rho}_{p}}
{2(2j_{b}+1)}} \int {\mid}N_{0}{\mid}^{2}d{\Omega}_{p}, \label{sl} \\
S_{T}(q,{\omega})&=&\sum_{{\mu}_{b},\, s_{p}}{\frac {{\rho}_{p}}
{2(2j_{b}+1)}} \int
({\mid}N_{x}{\mid}^{2}+{\mid}N_{y}{\mid}^{2})d{\Omega}_{p}, \label{st}
\end{eqnarray}
where the density of states for the outgoing nucleon is defined as ${\rho}_{p}={\frac {pE_{p}} {(2\pi)^{3}}}$.
The ${\hat {\bf z}}$-axis is taken to be along the three momentum transfer ${\bf q}$ and  the z-components of the angular momentum of the bound and continuum state nucleons are ${\mu}_{b}$ and $s_{p}$, respectively.
The Fourier transform of the nucleon transition current $J^{\mu}({\bf r})$ is simply given by
\begin{equation}
N^{\mu}=\int J^{\mu}({\bf r})e^{i{\bf q}{\cdot}{\bf r}}d^{3}r . \label{fourier}
\end{equation}
The continuity equation could be used to eliminate the $z$-component ($N_{z}$) via the equation $N_{z}=-{\frac {\omega} {q}}N_{0}$ if the current is conserved by using the relation $q_{\mu} N^{\mu}=0$.

The nucleon transition current in the relativistic single particle model is given by
\begin{equation}
J^{\mu}({\bf r})=e{\bar{\psi}}_{p}({\bf r}){\hat {\bf J}}^{\mu}
{\psi}_{b}({\bf r}) \;,
\end{equation}
where ${\hat {\bf J}}^{\mu}$ is a free nucleon current operator, and $\psi_{p}$ and $\psi_{b}$ are the wave functions of the outgoing nucleon and the bound state, respectively.
For a free nucleon, the operator comprises the Dirac contribution and the contribution of an anomalous magnetic moment  $\mu_{T}$ given by 
${\hat {\bf J}}^{\mu}=F_{1}(q_{\mu}^2){\gamma}^{\mu}+ i F_{2}(q_{\mu}^2){\frac{{\mu}_{T}} {2M}}{\sigma}^{\mu\nu}q_{\nu}$
with $q^2_\mu = - Q^2 = \omega^2 - {\bf q}^2$.
The form factors $F_{1}$ and $F_{2}$  are related to the electric and magnetic Sachs form factors given by 
$G_{E}=F_{1}+{\frac{{\mu}_{T}}{4M^{2}}}Q^2 F_{2}$ and 
$G_{M}=F_{1}+{\mu}_{T}F_{2}$ which are assumed to take the following standard form:
\begin{equation}
G_{E}={\frac {1} {(1 + {\frac {Q^2}{\Lambda^2})^{2}}}}
={\frac {G_{M}} {({\mu}_{T}+1)}} \;, \label{electric}
\end{equation}
where the standard value for $\Lambda^2$ is 0.71 (GeV/c)$^2$
and the anomalous magnetic moment $\mu_T$ is used 1.793 for proton and $-1.91$ for neutron.

\section{Result}

\subsection{Cross sections from SLAC data}

\begin{figure}
\begin{center}
\includegraphics[width=1.0\textwidth]{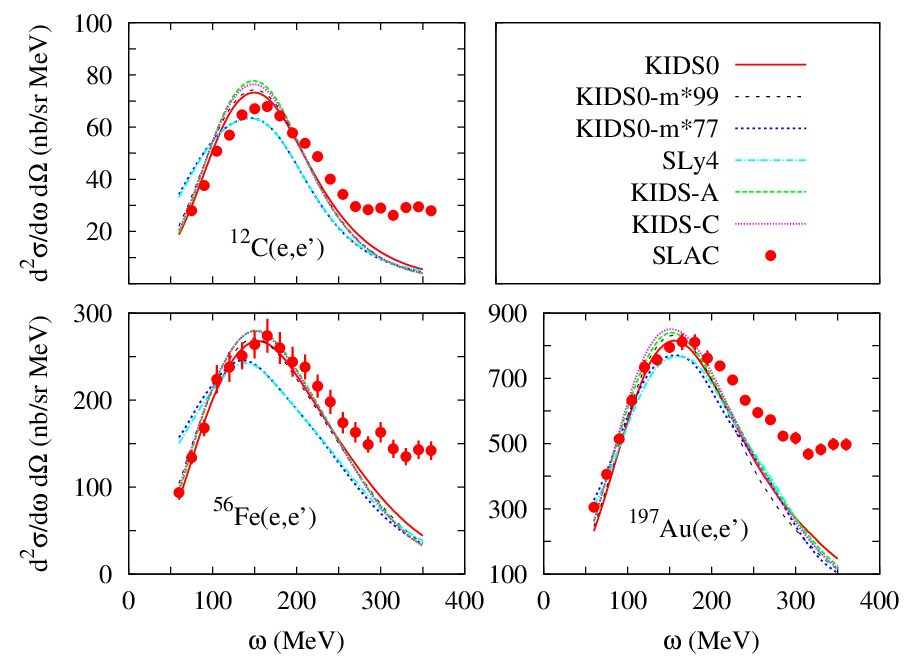}
\end{center}
\caption{Differential cross section for the incident electron energy 2.02 GeV and at the scattering angle $\theta = 15^\circ$.
SLAC data are from \cite{slacdata}.}
\label{slac} 
\end{figure}

\begin{figure}
\begin{center}
\includegraphics[width=1.0\textwidth]{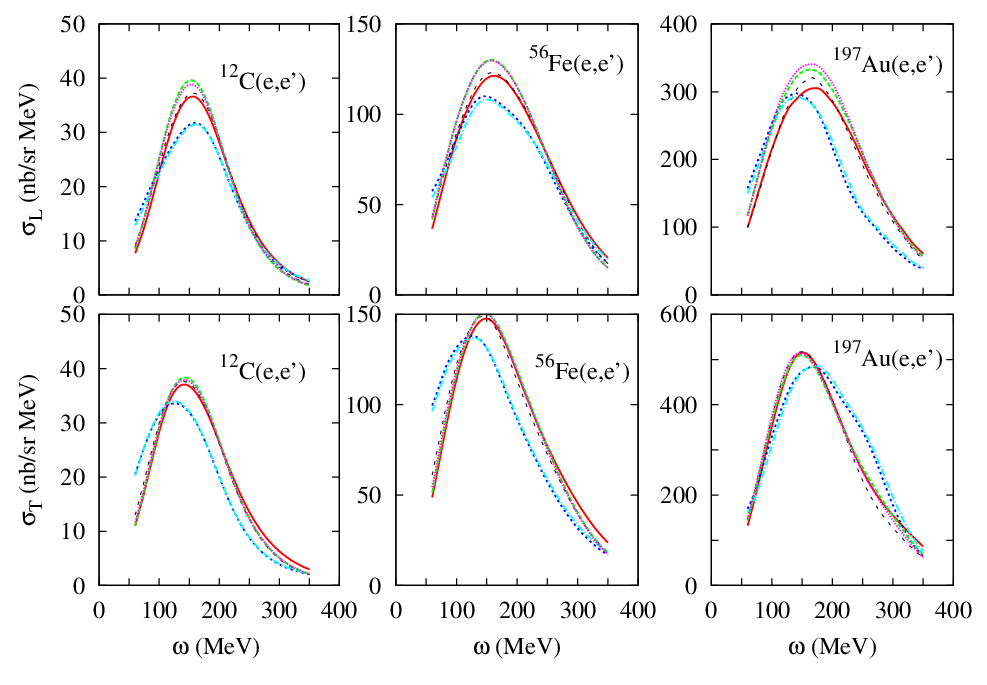}
\end{center}
\caption{Longitudinal and transverse cross sections for the incident electron energy 2.02 GeV and at the scattering angle $\theta = 15^\circ$.
The upper and lower panels are the results for the longitudinal and transverse cross sections, respectively.}
\label{slac-slst}
\end{figure}

SLAC data provide the differential cross sections with $^{12}$C, $^{56}$Fe, and $^{197}$Au for the incident electron energy 2.02~GeV, and at the scattering angle $\theta=15^\circ$ \cite{slacdata}.
The value of the four momentum transfer square is about $Q^2=0.25$ (GeV/c)$^2$ around the peak.
Figure \ref{slac} shows the data and theoretical results.
There are several features insensitive to the mass of nuclei. 
KIDS0-m*77 and SLy4 models behave almost the same, and they are worst in the agreement with experiment.
They are always underestimated than the other models KIDS0, KIDS0-m*99, KIDS-A and C.
KIDS0 and KIDS0-m*99 models are best at reproducing the data.
KIDS-A, C models predict the cross section larger than the KIDS0 and KIDS0-m*99 models, but the difference
is not as significant as the difference from the KIDS0-m*77 and SLy4 models.

One may have already noticed easily that the two groups, one with KIDS0-m*77 and SLy4 (GroupI),
and the other with the remaining 4 models (Group II) are distinguished in terms of the effective mass.
Group I models have the effective mass in the range $(0.7 \sim 0.8) M$, while the Group II models have
isoscalar effective masses $(0.9 \sim 1.0)M$ and isovector ones $(0.8 \sim 0.9)M$.
The results indicates tha large effective mass is favored over the effective masses smaller than $0.8M$ in the comparison with data.
A similar result has also been obtained in the exclusive $(e,e')$ reaction \cite{exeA2021}.
In the scattering with $^{208}$Pb, results for the 3s$_{1/2}$ and 2d$_{3/2}$ states show evident dependence on the effective mass
and better agreement to data with larger effective masses.
Response functions in the 1p$_{1/2}$ and 1p$_{3/2}$ states of $^{16}$O also show clear and better agreement to data with
KIDS0 and KIDS0-m*99 models than the KIDS-m*77 and SLy4.

Effect of the symmetry energy can be extracted by comparing KIDS0, KIDS-A and KIDS-C.
Around peaks ($\omega \sim 150$ MeV), difference within the three models is 5\% at most,
and the difference decreases as the mass number increases.
Three models have distinct values of $K_0$, $J$, $L$, and $K_\tau$, but the result of KIDS-A is very similar to that of KIDS-C.
These similarities demonstrate that the cross section is insensitive to the density dependence of the symmetry energy
at least within the range that nuclear data and neutron star properties are reproduced accurately.

It is worthwhile to compare the result of KIDS0 and SLy4.
Both models are partially fitted to the pure neutron matter EoS of \cite{apr}, so they have similar values of $J$ and $L$,
which determine the behavior of the symmetry energy most dominantly at densities relevant to nuclei.
Even though the symmetry energy is similar in the two models, cross sections are very different.
This comparison confirms that the dependence on the symmetry energy is insignificant compared to that of the effective mass.

In Fig. \ref{slac-slst}, we show the longitudinal and transverse cross sections in Eq. (\ref{csr})
\begin{eqnarray} 
\sigma_L &=& 
\sigma_M  {\frac {Q^4} {q^4}} S_L (q,\omega),   \nonumber \\  
\sigma_T &=& 
\sigma_M   \left (\tan^2 {\frac {\theta_e} {2}} + {\frac {Q^2} {2q^2}} \right ) S_T (q,\omega).
\end{eqnarray}
The upper and lower panels are the results for the longitudinal and transverse cross sections, respectively.
The kinematics are the same as the ones in Fig. \ref{slac}.
The difference between Group I and Group II is shown clearly in all the results.
While in the case of transverse part the effects of effective masses and the symmetry energy are very small in Group II, 
in the longitudinal part, the role of the symmetry energy in Group II enhances the magnitude of the cross section.
Peak positions in Fig.~\ref{slac} are similar between Group I and Group II for $^{12}$Ca and $^{197}$Au,
but clearly distinguished for $^{56}$Fe.
The reason could be understood from the behavior $\sigma_L$ and $\sigma_T$.
Peak position of Group I is to the left of Group II by about 50 MeV for $\sigma_T$ of $^{12}$C,
$\sigma_L$ and $\sigma_T$ of $^{56}$Fe, and $\sigma_L$ of $^{197}$Au,
but for the case of $\sigma_L$ for $^{12}$C and $\sigma_T$ for $^{197}$Au, peaks of Group I are located
on the right side of the peaks of Group II.
Portion of the longitudinal part in the total differential cross in Eq.~(\ref{csr}) is about 45 \% due to very forward scattering angle.

\subsection{Cross sections from Bates and Saclay data}

\begin{figure}
\begin{center}
\includegraphics[width=1.0 \textwidth]{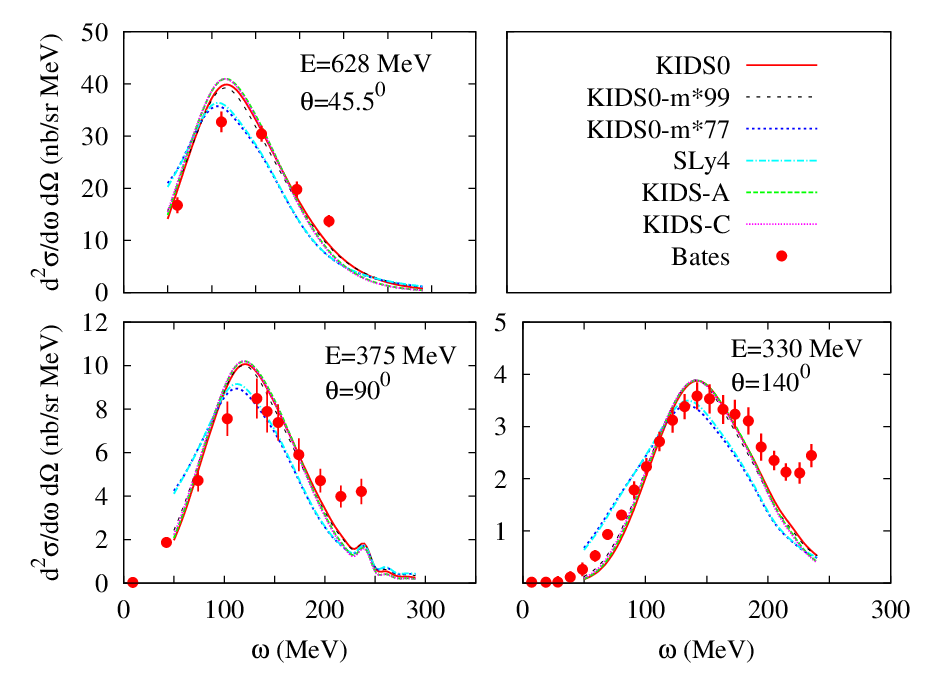}
\end{center}
\caption{Cross sections of the $^{40}$Ca($e$, $e'$) reaction. Experimental data are taken from \cite{bates}.}
\label{bates}
\end{figure}

Figure \ref{bates} shows the cross section of the $^{40}$Ca($e, e'$) reaction for the incident electron energies $E$ and 
at the scattering angle $\theta$ denoted in each panel.
The values of the four momentum transfer square around the peak are about $Q^2=0.22$ (GeV/c)$^2$ for 330 MeV 
and about $Q^2=0.19$ (GeV/c)$^2$ for 628 and 375 MeV.
Results are compared with experimental data from Bates \cite{bates}.

For $E=330$ MeV and $\theta = 140^\circ$, overall shape of data coincides better with Group II than Group I.
Position of the peak also agrees well with the models in Group II.
For $E=375$ MeV and $\theta = 90^\circ$, data around the peak agree well with Group I, and the Group II models predict
the cross sections slightly above the error bars.
Overall behavior before and after the peak is reproduced better by Group II.
For $E=628$ MeV and $\theta=45.5^\circ$, number of data is not sufficient to pin down the location of peak.
Both Group I and Group II predict similar position of peak.
For a few data below and above the peak, Group II models show good agreement with experiment.

\begin{figure}
\begin{center}
\includegraphics[width=1.0 \textwidth]{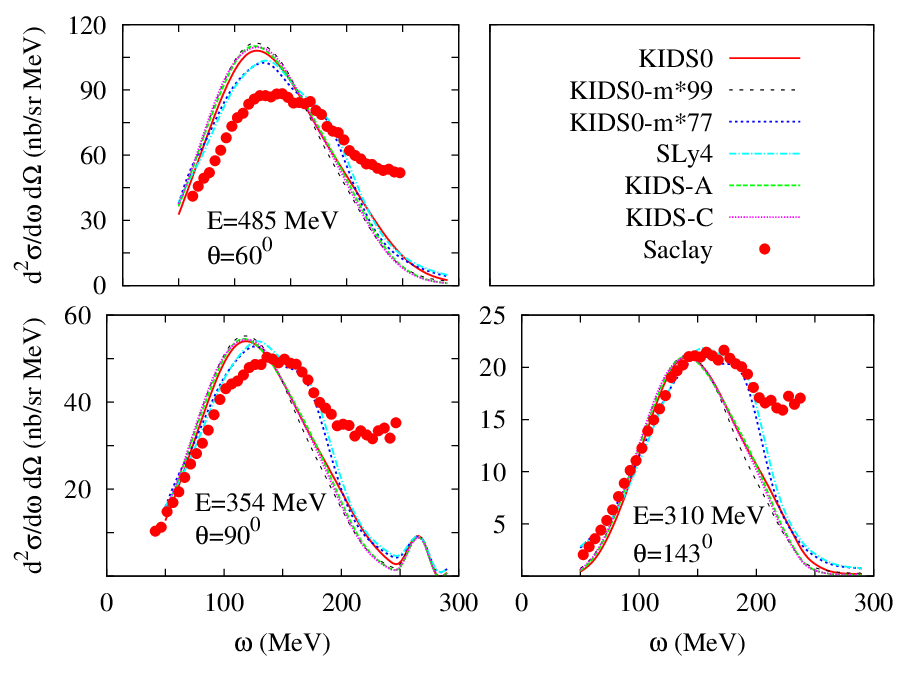}
\end{center}
\caption{Cross sections of the $^{208}$Pb($e$, $e'$) reaction. Experimental data are taken from \cite{saclay}.} 
\label{saclay}
\end{figure}

Figure \ref{saclay} shows the cross section of the $^{208}$Pb ($e, e'$) at the incident energies and the scattering angles specified in each panel.
Theoretical results are compared with experimental data taken from Saclay \cite{saclay}.
The values of the four momentum transfer square around the peak are about $Q^2=0.19$ (GeV/c)$^2$ for 310 MeV 
and about $Q^2=0.17$ (GeV/c)$^2$ for 485 and 354  MeV.
Agreement to data depends heavily on the energy.

For $E=310$ MeV and $\theta =143^\circ$,
all the models reproduce data up to the peak, and the difference is negligible between the models.
Shape of the peak is flat and wide over the energy transfer $\omega = 150 \sim 200$ MeV.
Such a behavior is unusual compared to the data in the reactions with $^{12}$C and $^{56}$Fe,
but the Group I models reproduce the data almost perfectly in this energy transfer range.
On the other hand, the Group II models go down quickly and monotonically right after the peak.
For $E=354$ MeV and $\theta = 90^\circ$, models predict similar result and agree with data
up to $\omega = 100$ MeV.
Group II models do not agree with the data at all above 100 MeV, but the Group I models reproduce the flat and wide
behavior in $\omega = 130 \sim 170$ MeV.
For $E=485$ MeV and $\theta=60^\circ$, Group I models give the result agreeing with data in $\omega =170 \sim 190$ MeV.
At the other transfer energies, theory fails to reproduce the data.

\begin{figure}
\begin{center}
\includegraphics[width=1.0\textwidth]{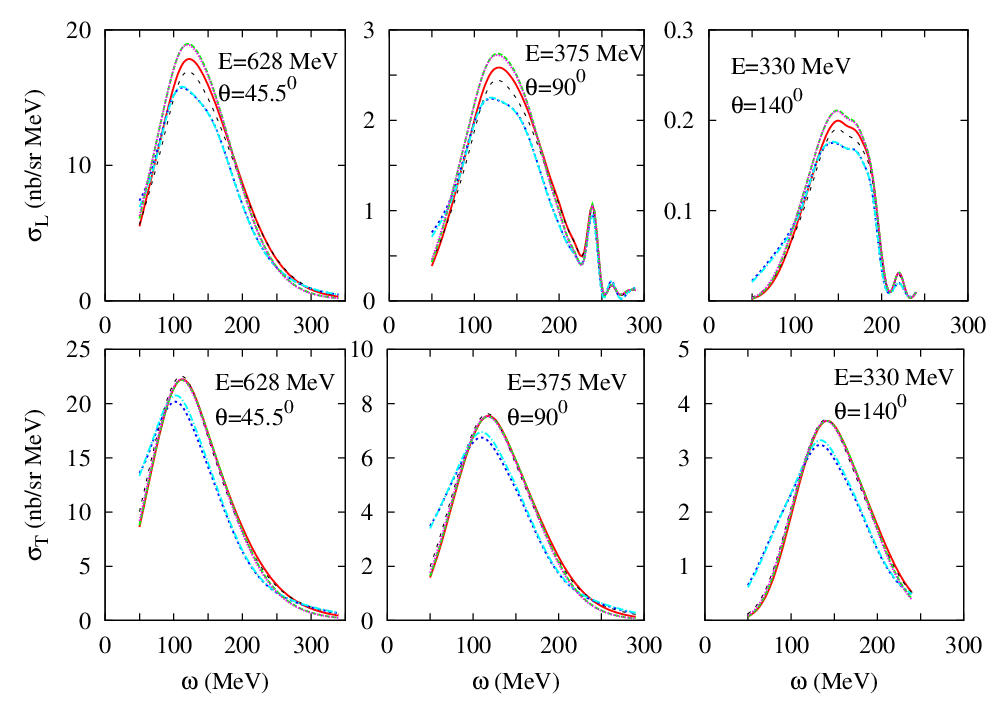}
\end{center}
\caption{Longitudinal and transverse cross section for $^{40}$Ca. The explanations for the curves are the same as those in Fig. \ref{bates}.} 
\label{ca-slst}
\end{figure}

\begin{figure}
\begin{center}
\includegraphics[width=1.0\textwidth]{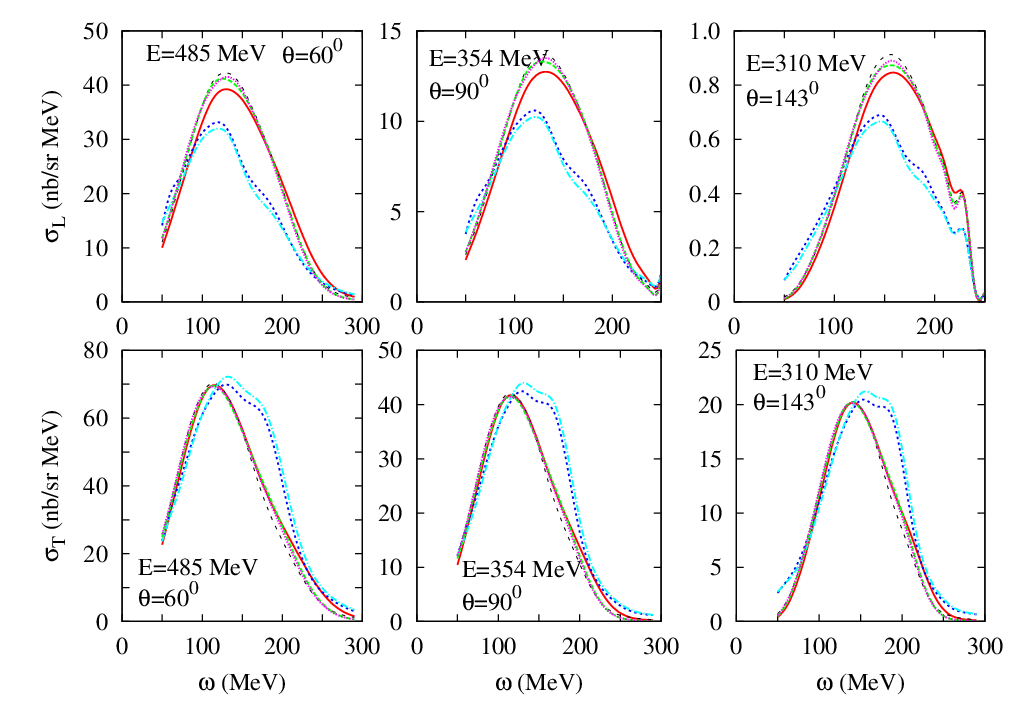}
\end{center}
\caption{Longitudinal and transverse cross section for $^{208}$Pb. The explanations for the curves are the same as those in Fig. \ref{saclay}.} 
\label{pb-slst}
\end{figure}

In Figs.~\ref{ca-slst} and \ref{pb-slst}, the longitudinal and transverse cross sections are shown for the reactions 
of $^{40}$Ca and $^{208}$Pb, respectively.
For the longitudinal cross sections, similar to Fig. \ref{slac-slst}, dependence on the symmetry energy is transparent,
and the peak positions of Group II is shifted to the right side of Group I by about 20 $\sim$ 30 MeV.
The differences for the peak positions decrease with larger scattering angle between Group I and II.
The reason for the suppression of $\sigma_L$ is because the term $\tan^2 \theta_e/2$ in $\sigma_T$,
cross section is dominated by $\sigma_T$ as $\theta_e$ increases.
In the transverse cross section, the influence of the effective mass and symmetry energy is seldom distinguishable within each Group.
Flat and wide peaks in Pb are originated from the behavior of $\sigma_T$ in the Group I models.
Group II models do not show such a behavior at all.
Interestingly, flat peaks also appear in $\sigma_L$ of Ca for the Group I models.
Flat shape becomes evident in $\sigma_L$ for Ca at $\theta_e = 140^\circ$.
However, because of the suppression by $\sigma_T$, the effect is not observed in Fig.~\ref{bates}.

Summarizing the results for the electron energies below 1 GeV,
it is hard to conclude definitively what values of the effective mass are more desirable for the consistency with experiment.
However, it is evidently certain that irrespective of the target nuclei, incident energy and scattering angle,
models in a group predict very similar results, and the results of Group I are manifestly
discriminated from those of Group II.

\section{Summary}

In the present work, we calculated the cross sections for inclusive $(e,e')$ reaction from 
$^{12}$C, $^{40}$Ca, $^{56}$Fe, $^{197}$Au, and $^{208}$Pb in the quasielastic region by using the KIDS nuclear density functional model.
To conserve the nuclear current and guarantee gauge invariance, wave functions of the bound- and continuum-state nucleons 
are generated by solving the Dirac equation with the scalar and vector potentials identical in the intital and final states.
Scalar and vector potentials in the Dirac equation are obtained from the transformation of non-relativistic potentials
that are determined from well-defined properties of finite nuclei and infinite nuclear matter, 
and there is no artificial adjustment of the potential to the scattering data.

The precise value of the effective mass has not been determined in experiment or theory.
By using the KIDS density functional formalism, effective mass values can be fixed to assumed values independent of the input data.
Consequently, they are not coupled to static properties of nuclei and EoS of nuclear matter. 
Inclusive scattering supports the isoscalar effective mass in the range $(0.9 \sim 1.0)M$.
In particular, the effect of the masses in the range $(0.7 \sim 0.9)M$ moves the peak positions and suppresses the magnitude of cross sections.

Uncertainty arising from the symmetry energy was also probed.
Density dependence of the symmetry energy was adjusted to both nuclear data and neutron star observation.
Even though the coefficients of the symmetry energy vary over a wide range, 
the effect of the symmetry energy is weak and makes only a few percent difference in the cross section.
However the effect of the symmetry energy appears clearly in the longitudinal cross section
though it is very small compared to the transverse cross section.

In conclusion, the inclusive $(e,e')$ reaction suggests criterion to pin down the nucleon's effective mass in nuclear medium.
Effect of the symmetry energy is not as manifest as the effective mass, 
but longitudinal cross sections, if measured precisely, could provide complementary constraints 
to the density dependence of the symmetry energy.

\section*{Acknowledgments}
This work was supported by the National Research Foundation of Korea (NRF) grant funded by the Korea government (No. 2018R1A5A1025563 and No. 2020R1F1A1052495).

\end{document}